\documentclass[]{aastex62}
\usepackage{xcolor}

\submitjournal{ApJ Letters}

\definecolor{maroon}{cmyk}{0, 0.87, 0.68, 0.32}

\shorttitle{M-star astrospheres and the role of GCRs within}
\shortauthors{Herbst et al.}

\begin{document}

% ASS22066R2
% accepted on 17.06.2020

\title{ON THE DIVERSITY OF M-STAR ASTROSPHERES\\
AND THE ROLE OF GALACTIC COSMIC RAYS WITHIN}

\correspondingauthor{Konstantin Herbst}
\email{herbst@physik.uni-kiel.de}

\author[0000-0001-5622-4829]{Konstantin Herbst}
\affil{Institut f\"ur Experimentelle und Angewandte Physik, Christan-Albrechts-Universit\"at zu Kiel, Leibnizstr. 11, 24118 Kiel, Germany}

\author[0000-0002-9530-1396]{Klaus Scherer}
\affiliation{Institut f\"ur Theoretische Physik IV, Ruhr-Universit\"at Bochum, Germany}

\author[0000-0001-8264-1044]{Stefan E. S. Ferreira}
\affiliation{Centre for Space Research, North-West University, 2520 Potchefstroom, South Africa}

\author[0000-0002-4192-2082]{Lennart R. Baalmann}
\affiliation{Institut f\"ur Theoretische Physik IV, Ruhr-Universit\"at Bochum, Germany}

\author[0000-0003-3659-7956]{N. Eugene Engelbrecht}
\affiliation{Centre for Space Research, North-West University, 2520 Potchefstroom, South Africa}

\author[0000-0002-9151-5127]{Horst Fichtner}
\affiliation{Institut f\"ur Theoretische Physik IV, Ruhr-Universit\"at Bochum, Germany}

\author[0000-0001-6122-9376]{Jens Kleimann}
\affiliation{Institut f\"ur Theoretische Physik IV, Ruhr-Universit\"at Bochum, Germany}

\author[0000-0002-0205-0808]{R. Du Toit Strauss}
\affiliation{Centre for Space Research, North-West University, 2520 Potchefstroom, South Africa}

\author{Daniel M. Moeketsi}
\affiliation{Centre for High Performance Computing (CHPC), National Integrated Cyber-Infrastructure System (NICIS) - CSIR, Cape Town, South Africa}
\affiliation{Centre for Space Research, North-West University, 2520 Potchefstroom, South Africa}

\author[0000-0002-1856-9225]{Shazrene Mohamed}
\affiliation{Department of Astronomy, University of Cape Town, South Africa}

\begin{abstract}
%
% 245/250 words
%
With upcoming missions such as the James Webb Space Telescope (JWST), the European Extremely Large Telescope (ELT), and the Atmospheric Remote-sensing Infrared Exoplanet Large-survey (ARIEL), we soon will be on the verge of detecting and characterizing Earth-like exoplanetary atmospheres for the first time. These planets are most likely to be found around smaller and cooler K- and M-type stars. However, recent observations showed that their radiation environment might be much harsher than that of the Sun. Thus, the exoplanets are most likely exposed to an enhanced stellar radiation environment, which could affect their habitability, for example, in the form of a hazardous flux of energetic particles. Knowing the stellar radiation field, and being able to model the radiation exposure on the surface of a planet is crucial to assess its habitability. In this study, we present 3D magnetohydrodynamic (MHD)-based model efforts investigating M-stars, focusing on V374 Peg, Proxima Centauri, and LHS 1140, chosen because of their diverse astrospheric quantities.
We show that V374 Peg has a much larger astrosphere (ASP) than our Sun, while Proxima Centauri and LHS 1140 most likely have ASPs comparable or even much smaller than the heliosphere, respectively. Based on a 1D transport model, for the first time, we provide numerical estimates of the modulation of Galactic cosmic rays (GCRs) within the three ASPs. We show that the impact of GCRs on the Earth-like exoplanets Proxima Centauri b and LHS 1140 b cannot be neglected in the context of exoplanetary habitability.
\end{abstract}


%-------------------------------------------------------------------------------------------
\section{Introduction} \label{sec:intro}
%
% 442 words
%
The structure of an astrosphere (ASP) strongly depends on the properties of both the hot and fully ionized stellar wind and the surrounding interstellar medium (ISM). Differences in the stellar wind and the local ISM can lead to a wide range of shapes of ASPs \citep[e.g.,][]{Mueller-etal-2006, Scherer-etal-2008}. While simulations of ASPs around hot stars most often make use of 1D or 2D {\rm \mbox{(M)}MH} approaches \citep[e.g.,][]{vanMarle-etal-2014}, it is inevitable to apply 3D astrospheric modeling in order to estimate the radiation and cosmic-ray (CR) particle field of cool stars. In order to do so, for example, the 3D MHD code CRONOS \citep{Kissmann-etal-2018} can be used. CRONOS has been successfully applied to hot O-B stars \citep[e.g.,][]{Scherer-etal-2015} and, most recently, to a study of astrospheric shock structures \citep{Scherer-etal-2020}. 

While orbiting its parent star, potential close-in exoplanets of M-stars might be exposed to a strong outflow of stellar plasma, determining not only the planetary particle environment but also its magnetospheric properties \citep[see, e.g.,][]{Preusse-etal-2005}. From the Sun, we further know that intense solar flares often are accompanied by coronal mass ejections (CMEs). Keeping in mind that active stars produce much stronger and more frequent stellar flares, a substantial mass loss from flare-associated CMEs may be expected. Thus, long-term exposure to strong ambient stellar winds and CMEs may have substantial effects on planetary atmospheres through, for example, erosion where planets will lose a significant fraction of their atmosphere. Although \citet{Moschou-etal-2019} presented a more optimistic scenario of lower resulting CME kinetic energies, in order to sustain an atmosphere in closer-in habitable zones (HZs), planets would either need strong internal magnetic fields or thick atmospheres \citep[e.g., ][]{Lammer-etal-2007}. First studies have been performed, for example, by \citet{Khodachenko-etal-2007}. However, at this point, the impact of stellar winds and CMEs on close-in exoplanets is still controversially debated. 

Most recently the influence of CRs on atmospheric chemistry and climate came into focus \citep[e.g.,][]{Griessmeier-etal-2016, Scheucher-etal-2018, Herbst-etal-2019b}. In particular, the propagation of Galactic cosmic rays (GCRs) is highly influenced by the presence of astrospheric magnetic fields (AMFs), which act as small-scale sinks, decreasing the GCR flux. However, up to now, only analytic estimates of the GCR flux within other ASPs, in particular, those of M-stars, can be found in the literature \citep[e.g.,][]{Sadovski-etal-2018}. According to these estimates, the influence of GCRs can be neglected when studying exoplanetary habitability. 

Utilizing full 3D MHD modeling, we study the ASPs of the three M-stars V374 Pegasi, Proxima Centauri, and LHS 1140. In order to give a numerical estimate of the location- and energy-dependent CR flux within these ASPs, we further solve the particle transport equation \citep{Parker-1965} in a first approximation using 1D stochastic differential equations \citep[SDEs; see, e.g.,][]{Strauss-etal-2017}. 

%----------------------------------------------------------------------------------------------------------------------------
\subsection{Characteristics of M-stars}
%
% 346 words
%
Cool, low-mass stars like K- and M-dwarfs are more common within the Galaxy than hotter, more massive ones \citep[][]{Smith-Scalo-2009}. Their large number, long main-sequence lifetime, and their low luminosity (which is caused by their low masses and small radii) make them favorable targets to detect habitable rocky (Earth-like) exoplanets \citep[][]{Dittmann-etal-2017}. However, such cool stars have close-in HZs, regions where planetary temperatures are just right to sustain liquid water on the planetary surface, ranging between 0.03~au (late-type M-dwarfs) and 0.5~au (young M-dwarfs). According to \citet{West-etal-2004, West-etal-2015} and \citet{Mohanty-etal-2002} the more prominent the stellar convection envelope and the stellar rotation rates, the stronger the stellar activity. However, for mid- to late-type M-stars (M4 to M8.5 types), the activity saturates at higher rotational velocities, while above M9, the activity levels decrease significantly \citep[see, e.g.,][]{Kay-etal-2016}. Furthermore, according to \citet{Vidotto-etal-2011}, observations of surface magnetic field distributions suggest that young M-dwarfs host weak large-scale magnetic fields dominated by toroidal and non-axisymmetric poloidal configurations \citep[see also][]{Donati-etal-2006}. Mid-M-dwarfs, on the other hand, are hosts to strong, mainly axisymmetric large-scale poloidal fields \citep[][]{Morin-etal-2008}. According to \citet{Candelaresi-etal-2014}, for most of these stars, a stellar activity significantly above the solar level is observed.

Thus, the exoplanetary radiation environment around certain M-dwarfs may be much harsher as compared to that of the Sun \citep[][]{Herbst-etal-2019a} with which we are familiar. Due to their long main-sequence lifetimes and, therefore, activity periods in the order of Gyrs \citep{West-etal-2004}, as well as the small planet-star separations, exoplanets could be exposed to an enhanced stellar radiation field over long timescales. This radiation field, in turn, could affect the planetary habitability, for example, due to a hazardous flux of stellar energetic particles (SEPs) influencing its atmospheric evolution, climate, and photochemistry \citep[e.g.,][]{Scheucher-etal-2018, Scheucher-etal-2020a} as well as the altitude-dependent atmospheric radiation dose \citep[e.g.,][]{Atri-2020}.

Thus, detailed knowledge of the stellar radiation and particle environment and their impact on the (exo)planetary atmospheric chemistry, climate, and induced atmospheric particle radiation field is crucial in order to assess its habitability and, in particular, potential atmospheric biosignatures. However, up to now, the impact of GCRs has been neglected in such studies.
\begin{table}[!t]
\caption{Stellar properties of V374 Peg, Proxima Centauri, and LHS  1140 (first block), the corresponding stellar wind properties (second block), the assumed ISM parameters (third block), and the modeled termination shock (TS), astropause (AP), and bow shock (BS) distances (fourth block).}  
\label{tab:1}
%\centering
\begin{tabular}{l  c c c}       
\hline\hline                      
Parameter & V374 Peg & Prox Cen & LHS1140 \\
\hline
Type & M3.5Ve$^{(a)}$ & M5.5Ve$^{(a)}$ & M4.5$^{(a)}$\\
$T_{\mathrm{eff}}$ [K] & 3440$^{(b)}$ & 3050$^{(c)}$ & 3131$^{(d)}$ \\
$P_{\mathrm{rot}}$ [d] & 0.44$^{(b)}$ & 83$^{(e)}$ & 82.6$^{(d)}$  \\
$L_\star/L_\sun$ & 0.01452$^{(f)}$ & 0.00155$^{(f)}$ & 0.00298$^{(f)}$\\
$R_\star/R_\sun$ & 0.340$^{(b)}$ & 0.141$^{(c)}$ & 0.186$^{(d)}$\\
$M_\star/M_\sun$ & 0.280$^{(b)}$ & 0.123$^{(c)}$ & 0.146$^{(d)}$\\
$B_\star$ [nT] & $1.6\cdot 10^{8(g)}$ & $6\cdot 10^{7(h)}$ & $1\cdot10^{7(i)}$\\
$\dot{M_\star}$ [M$_\odot$ yr$^{-1}$] & 4$\cdot 10^{-10(g)}$ & 2$\cdot 10^{-15(j)}$ & 5$\cdot 10^{-17(k)}$ \\
\hline
\hline
$T_{\mathrm{sw}}$ at 1\,au [K]& 1$\cdot 10^5$$^{(l)}$ & 1$\cdot 10^5$$^{(l)}$ & 4.6$\cdot 10^4$$^{(l)}$\\
$v_{\mathrm{sw}}$ at 1\,au [km s$^{-1}$]& 1500$^{(g)}$ & 1500$^{(m)}$ & 250$^{(n)}$\\
$n_{\mathrm{sw}}$ at 1\,au [cm$^{-3}$] & 35742$^{(k)}$ & 0.00028$^{(k)}$ & 0.27$^{(k)}$\\
$B_{\mathrm{sw}}$ at 1\,au [nT]& 6.6$^{(o)}$ & 1.8$^{(o)}$ & 0.3$^{(o)}$\\ \hline
\hline
$T_{\mathrm{ISM}}$ [K] & 9000$^{(l)}$ & 9000$^{(l)}$ & 9000$^{(l)}$ \\
$v_{\mathrm{ISM}}$ [km s$^{-1}$] & 30$^{(l)}$ & 30$^{(l)}$ & 40$^{(l)}$ \\
$n_{\mathrm{ISM}}$ [cm$^{-3}$] & 11$^{(l)}$ & 0.06$^{(l)}$ & 0.1$^{(l)}$ \\
$B_{\mathrm{ISM}}$ [nT] & 1$^{(l)}$ & 0.3$^{(l)}$ & 0.3$^{(l)}$ \\\hline
\hline
TS [au] &3.6$\cdot 10^3$ & 53.7 & 8.0\\
AP [au] & 8.5$\cdot 10^3$& 122.0 & 11.3\\
BS [au]  & 1.3$\cdot 10^4$ & -- & 28.9
\\\hline
\multicolumn{4}{p{\columnwidth}}{
$^{a}$\footnotesize{Data taken from \url{http://simbad.u-strasbg.fr/simbad/}, 
$^{b}$\footnotesize{taken from \citet{Vida-etal-2016}},
$^{c}$\footnotesize{taken from \citet{Anglada-Escude-etal-2016}},
$^{d}$\footnotesize{taken from \citet{Dittmann-etal-2017}},
$^{6}$\footnotesize{taken from \citet{Benedict-etal-1998}}, %\citet{Ribas-etal-2016}}
$^{f}$\footnotesize{data calculated via the Stefan-Boltzmann law $L_\star \propto R_\star^2 T_\star^4$}, 
$^{g}$\footnotesize{taken from \citet{Vidotto-etal-2011}}, 
$^{h}$\footnotesize{taken from \citet{Reiners-Basri-2008}}, %from \citet{Garraffo-etal-2016}},
$^{i}$\footnotesize{taken from \citet{Donati-etal-2006}}, 
$^{j}$\footnotesize{taken from \citet{Wood-etal-2001}}, 
$^{k}$\footnotesize{after \citet{Wilkin-2000} (see text)},
$^{l}$\footnotesize{sophisticated guess based on heliospheric parameters}, 
$^{m}$\footnotesize{within the limitations of \citet{Garraffo-etal-2016}},  
$^{n}$\footnotesize{scaling with the ratio of solar and  stellar temperature}}, 
$^{o}$\footnotesize{calculated after $B_{sw} = B_\star/B_\odot \cdot B_{\odot,sw}$}}\\
\end{tabular}
\end{table}
%----------------------------------------------------------------------------------------------------------------------------
\subsection{Characteristics of the studied systems}\label{sec:characteristics}
%
% 293 words
%
In order to investigate the diversity of M-star ASPs, in this study, the M-dwarfs V374 Pegasi (huge ASP compared to the heliosphere) and Proxima Centauri (medium-sized ASP comparable to the heliosphere), which both are known to be active flaring stars, and LHS 1140 (tiny ASP compared to the heliosphere), an inactive star, are studied. Their characteristic features, such as luminosity, radius, and mass, are listed in the upper part of Table~\ref{tab:1}. 
\subsubsection{V374 Pegasi}
V374 Pegasi, also-known as GJ 4247, is an old main-sequence star with an extraordinarily strong surface magnetic field of about $1.6\cdot 10^8$\,nT and a mass-loss rate of $4\cdot 10^{-11}$ to $4\cdot10^{-10}\,M_\sun$ yr$^{-1}$ \citep[][]{Vidotto-etal-2011}. 
\subsubsection{Proxima Centauri}
Our nearest neighbor Proxima Centauri, also known as GJ 551, is only $(1.3012 \pm 0.0003)$\,pc away \citep[][]{Garraffo-etal-2016}. According to \citet{Reiners-Basri-2008} and  \citet{Garraffo-etal-2016}, its surface magnetic field is in the order of $6\cdot 10^{7}$\,nT, and has a mass-loss rate of approximately $2.84 \cdot 10^{-15}\,M_\sun$ yr$^{-1}$. Its rocky Earth-like planet Proxima Centauri b has a semi major axis of $(0.0485 \pm 0.0041)$\,au, a mass of $m_{\mathrm{P}}$ = 1.63$^{+1.66}_{-0.72}\,m_\earth$ and a radius of $R_{\mathrm{P}} = 1.07^{+0.38}_{-0.31}\,R_\earth$ \citep[see][]{Bixel-Apai-2017}. With an equilibrium temperature of 227\,K, its orbital period is in the order of 11.186$^{+0.001}_{-0.002}$\,d.
\subsubsection{LHS 1140}
LHS 1140, also known as GJ 3053, can be found at a distance of $(12.47 \pm 0.42)$\,pc and is more than 5\,Gyrs old \citep{Dittmann-etal-2017}. Its rocky super-Earth, LHS 1140 b, whose orbit has a semi major axis of $(0.0875 \pm 0.0041)$\,au, has a mass of $m_{\mathrm{P}} = (6.65 \pm 1.82)\,m_\earth$, a radius of $R_{\mathrm{P}} = (1.43 \pm 0.10)\,R_\earth$, a surface gravity of $g_{\mathrm{P}} = (31.8 \pm 7.7)$\,m s$^{-2}$, and an equilibrium temperature of $T_{\mathrm{eq}} = (230 \pm 20)$\,K \citep{Dittmann-etal-2017}.
%-------------------------------------------------------------------------------------------
%\linebreak
\section{Modelling stellar astrospheres}
%
% 540 words
%
In order to model the ASPs of the three M-dwarfs, we use the 3D finite-volume MHD code \textsc{CRONOS} \citep{Kissmann-etal-2018}. The code is based on a Riemann solver in order to perform simulations on a star-centered spherical grid with a resolution of
$N_r \times N_\vartheta \times N_\varphi = 1024 \times 60 \times 120$ cells in the cases of LHS 1140, $1024 \times 64 \times 32$ cells in the case of Proxima Centauri, and $1024 \times 16 \times 32$ cells in the case of V374 Peg. Thereby, distances to the star within $[0.03\,\mathrm{au}...1\,\mathrm{pc}]$ and the full $4\pi$ solid angle are covered; here the approach by \citet{Scherer-etal-2020} is followed. 

We note that the stars have been chosen carefully based on the variety of their stellar magnetic fields: while V374 Peg has a magnetic field that is about three times stronger than the heliospheric magnetic field (HMF), the magnetic field of LHS1140 is about one order of magnitude weaker than the HMF. Unfortunately, no direct information regarding the ISM in the vicinity of V347 Peg and LHS 1140 are available, and certain assumptions have to be made. The assumed ISM parameters are listed in the third block of Table~\ref{tab:1}. 

In case the stellar magnetic field is known, only information of the magnetic field strength on the surface of the star is available. By assuming that the stellar magnetic field is an analog to the HMF and thus is frozen into the stellar wind and forms a Parker spiral \citep{Parker-1958}, the AMF can be assumed as
\begin{equation}\\
B=B_{0} \frac{r_{0}^{2}}{r^{2}} \sqrt{1+\left(\frac{r \Omega}{v_{\mathrm{sw}}}\right)^{2} \sin^{2}{\vartheta}} \ ,
\end{equation}
where $B_0$ is the radial magnetic field component at distance $r_0$ from the star and $r$ the stellar-centric distance. In a frame corotating with the star both the frozen-in field line and the stream line of the plasma coincide. Therefore, the ratio between the azimuthal stellar wind speed $v_{\varphi}$ and the radial stellar wind speed $v_{\mathrm{sw}}$ is given by 
\begin{equation}
\frac{v_\varphi}{v_{\mathrm{sw}}} = \frac{\Omega r \sin{\vartheta}}{v_{\mathrm{sw}}},
\end{equation}
where $\Omega$ is the stellar rotation rate and $\vartheta$ the colatitude.
\begin{center}
\begin{figure*}[!t]
 \includegraphics[width=\textwidth]{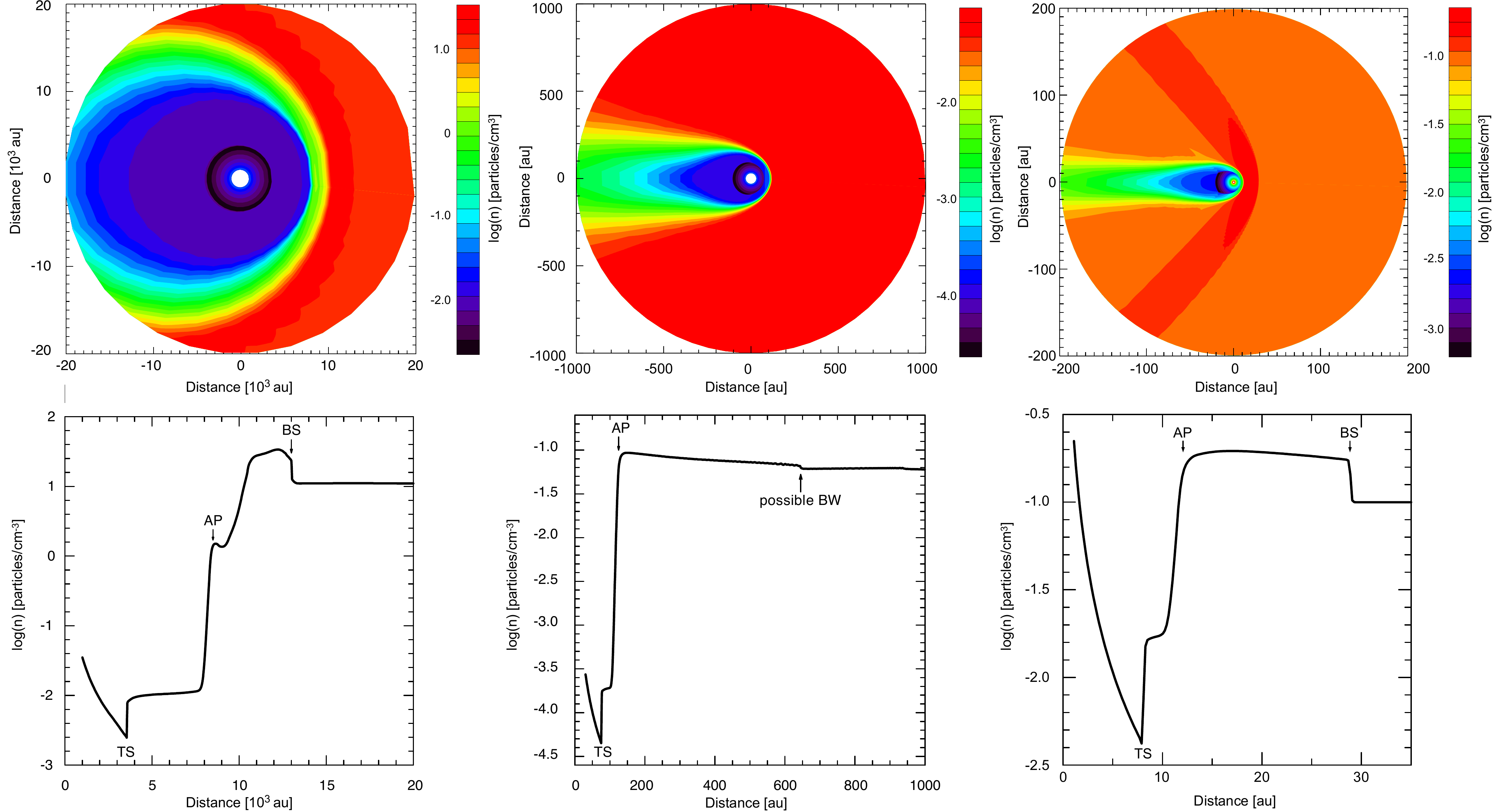}
  \caption{Upper panels: Modeled density distribution of V374 Peg (left), Proxima Centauri (middle), and LHS1140 (right) shown within the equatorial plane. In all cases, the ISM is flowing in from the right-hand side. Lower panels: The corresponding spatial density profiles along the line of sight toward the incoming ISM. Termination shock (TS), astropause (AP), and Bow shock (BS) distances are marked. We note that the presented results are based on the values given in Table~\ref{tab:1}.}
 \label{fig:1}
 \end{figure*}
\end{center}
Another critical parameter is the distance of the astrospheric termination shock (TS). The defining quantity, determining this distance, is the balance between the momentum density of the ISM and that of the supersonic stellar wind, both dominated by their ram pressures. According to \citet{Parker-1958}, the TS distance in the upwind direction is given by
\begin{equation}
r_{\mathrm{TS}}={r_{0} \frac{v_{\mathrm{sw}}}{v_{\mathrm{ISM}}}} \sqrt{\frac{\rho_{\mathrm{sw}}}{\rho_{\mathrm{ISM}}}},
\end{equation}
with $r_0$ the  reference distance, $v_{\mathrm{ISM}}$ the speed of the ISM while $\rho_{\mathrm{sw}}$ and $\rho_{\mathrm{ISM}}$ give the stellar wind and ISM densities, respectively. Furthermore, $\rho_{\mathrm{sw}}$ can be determined by taking into account the stellar mass loss rate $\dot{M_\star}$ \citep{Wilkin-2000}, leading to $\rho_{\mathrm{sw}}(r_{0})=\dot{M}_{\star}/(4 \pi r_0^{2} v_{\mathrm{sw}})$, with $r_0$ = 1 au, and thus
\begin{equation}
r_{\mathrm{TS}}=\sqrt{\frac{\dot{M_\star} v_{\mathrm{sw}}}{4 \pi \rho_{\mathrm{ISM}} v_{\mathrm{ISM}}^{2}}}.
\end{equation}

We note that due to the lack of observational information, the stellar wind temperatures of the three M-stars at 1 au have been assumed to be comparable to those of the solar wind. Due to the supersonic character of the stellar wind, however, the thermal stellar wind pressure (and thus the assumed stellar wind temperature) does not play an important role. As discussed in \citet{Scherer-etal-2020}, in highly supersonic flows, the ram pressure is much larger than the thermal pressure, and about three-quarters of the ram pressure is converted to thermal pressure at the termination shock.

\section{Results and Discussion}
%
% 334
%
\subsection{The astrospheric structure of V374 Peg, Proxima Centauri, and LHS1140}\label{sec:astrospheres}
Figure~\ref{fig:1} shows the density distribution according to our 3D MHD model effort to determine and characterize the ASPs of V374 Peg (left panel), Proxima Centauri (middle panel), and LHS 1140 (right panel). Significant differences between the ASPs of the modeled M-stars are evident. For example, for V374 Peg, with an ASP expanding up to 13'000 au due to, amongst other processes, its high mass-loss rate of about $4\cdot 10^{-10} M_\sun$ yr$^{-1}$ (see Table~\ref{tab:1}), the stellar flow dominates the ISM flow, resulting in the formation of a cavity that is much larger than the heliosphere while the inner AMF is almost negligible. 

LHS 1140, on the other hand, has a surprisingly small ASP with a (line-of-sight) TS at about 8.1\,au, an astropause (AP) at around 11.5\,au, and a bow shock (BS) at $\sim$ 28.9~au. Thus, contrary to common assumptions, our model efforts show that not all cool M-stars drive huge ASPs that protect potential Earth-like rocky exoplanets sheltered within from GCRs accelerated at supernovae remnants to energies of up to hundreds of TeV \citep[e.g.,][]{Buesching-etal-2005}. 

\begin{figure}[!t]
\begin{center}
 \includegraphics[width=0.6\columnwidth]{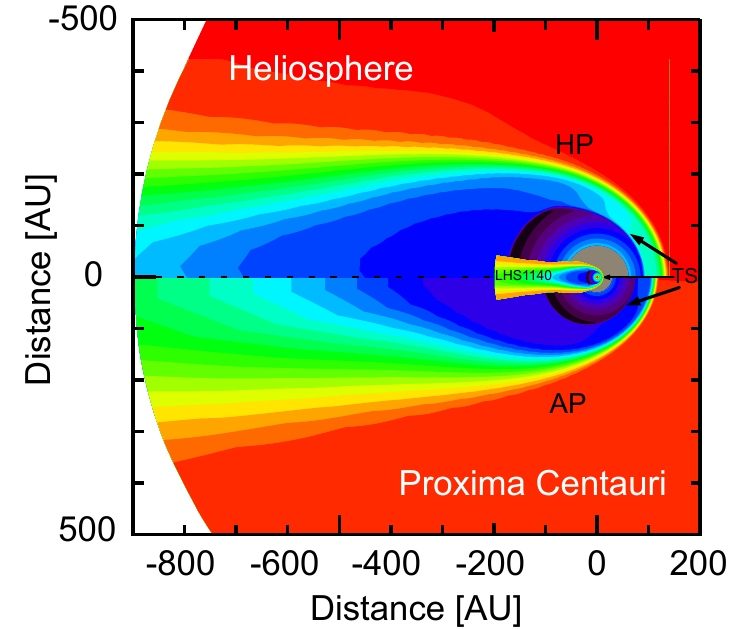}
  \caption{Direct comparison of the model results for the heliosphere (upper part), Proxima Centauri (lower part), and LHS 1140 \textbf{(on top)}. The locations of the TS and the HP/AP are highlighted. Note that the modeled heliospheric TS and HP distances are located at 90 au and 130 au, respectively.}
 \label{fig:2}
 \end{center}
 \end{figure}

In particular, the ASP of Proxima Centauri, with its (line-of-sight) TS at 76\,au, and its AP at 110\,au, is surprisingly similar to the heliosphere. A direct comparison of both ASPs is given in Fig.~\ref{fig:2}. Here, the upper half shows the model results for the heliosphere, while the lower half highlights the model results for Proxima Centauri. Proxima Centauri has a less expanded TS in the direction towards the incoming ISM than the heliosphere, and its AP is much closer. Furthermore, Proxima Centauri has a more compressed tail-ward TS compared to the heliosphere. On top, also the ASP of LHS1140 is shown to scale.

Moreover, Proxima Centauri is found not to drive a BS, which most likely will change when applying a multi-fluid approach. As discussed by, for example, \citet{Scherer-etal-2014}, including the He$^+$ component will result in astrospheric alfv\'{e}nic and magnetosonic wave speeds lower than the flow speed of the LISM, and lead to the build-up of a BS \cite[see also][]{Izmodenov-Alexashov-2015}.
\begin{table*}[!t]
\label{tab:2}
\caption{Energy-dependent differential GCR intensities at Earth, Proxima Centauri b, and LHS 1140 b in units of m$^{-2}$ sr$^{-1}$ s$^{-1}$ GeV$^{-1}$. Upper values correspond to stellar minumum- and lower values to stellar maximum conditions.}  
\label{tab:2}
\centering
\begin{tabular}{l  c  c  c  c  c  c }       
\hline\hline                      
Energy&  0.1 GeV & 0.4 GeV & 1.0 GeV & 4.0 GeV & 10.0 GeV & 40.0 GeV\\
\hline
\hline
LIS  & 26.994 & 15.37 & 5.58 & 0.33 & 2.97$\cdot 10^{-2}$ & 6.31 $\cdot 10^{-4}$\\
\hline
\hline
LHS 1140 b & 20.31 & 11.75 & 4.80 & 0.26 & 2.60$\cdot 10^{-2}$ & 6.00$\cdot 10^{-4}$\\
  &     18.15 & 9.67  & 3.60 & 0.18 & 2.00$\cdot 10^{-2}$ & 5.30$\cdot 10^{-4}$\\
\hline
\hline
Prox Cen b   & 19.29 & 10.23 & 3.92 & 0.21 & 2.21$\cdot 10^{-2}$ & 5.84$\cdot 10^{-4}$\\
      & 16.59 & 7.88 & 2.77 & 0.15 & 1.79$\cdot 10^{-2}$ & 4.84$\cdot 10^{-4}$\\
\hline
\hline
Earth  &1.16 &1.93 &1.35& 0.22& 2.0$\cdot 10^{-2}$ & 5.5 $\cdot 10^{-4}$\\
&  0.27&0.64&0.63&0.14&2.0$\cdot 10^{-2}$&5.5 $\cdot 10^{-4}$\\
\hline
\hline
V374 Peg  &4.10$\cdot 10^{-2}$&	4.02$\cdot 10^{-2}$&	8.90$\cdot 10^{-3}$&	5.10$\cdot 10^{-3}$&	2.37$\cdot 10^{-3}$&	3.87$\cdot 10^{-4}$\\
&  9.80$\cdot 10^{-5}$ &	1.30$\cdot 10^{-4}$&	2.00$\cdot 10^{-4}$&	1.40$\cdot 10^{-4}$&	1.10$\cdot 10^{-4}$&	7.90$\cdot 10^{-5}$\\
\hline
\hline
\end{tabular}
\end{table*}
\subsection{Modulation of GCRs inside the astrospheres of V374 Peg, Proxima Centauri, and LHS 1140}
%
% 1076
%
From the heliosphere, we know that the turbulent HMF influences the energy spectrum of GCRs by modulating the spectrum below $\sim 40$~GeV. Nevertheless,  not only SEPs but also GCRs play an essential role within, for example, the CO$_2$- or N$_2$-O$_2$-dominated atmospheres of Earth, Venus, and Mars, respectively. Particularly significant are the induced changes of the atmospheric ionization, and thus the atmospheric chemistry, as well as the atmospheric radiation dose, which is a measure for planetary habitability. 

Based on an analytic approach, \citet{Sadovski-etal-2018}, for example, found that GCRs below 1~TeV are not present at the orbit of {Proxima Centauri b}. The same results are achieved when the so-called force-field approach \citep[e.g.,][]{Caballero-Lopez-Moraal-2004} is applied. Thus, in theory, GCRs can be neglected when it comes to studying the impact of CRs on exoplanetary atmospheres.  However, the latter employs an AMF much stronger than the HMF, which, according to our model results, is not a valid assumption for all M-star ASPs. 

Therefore, as a first step, we utilize the stellar wind speed and magnetic field distributions along the stagnation line provided by our 3D MHD modeling efforts in order to numerically investigate the modulation of GCRs within the ASPs of {V374~Peg}, {Proxima Centauri}, and {LHS 1140}. Numerical models to determine the modulation level of GCRs within the heliosphere are based on solving the transport equation of \citet{Parker-1965}. As already mentioned in \citet{Caballero-Lopez-Moraal-2004}, it is advisable to use a full 1D transport code. Thus, in this study, a 1D version of the transport equation is solved numerically by means of SDEs \citep[][and references therein]{ Strauss-etal-2017} in order to investigate the importance of GCRs within the ASPs of M-stars for the first time. 

Solving the transport equation in radial direction leads to a radial diffusion coefficient 
\begin{equation}
\kappa_{rr} = \kappa_{\perp r} \sin^2{\Psi} + \kappa_{\parallel} \cos^2{\Psi},
\end{equation}
where $\Psi$ is the heliospheric/astrospheric winding angle. However, because of the stellar rotation, a largely azimuthal magnetic field is present in ASPs, leading to $\Psi \rightarrow 90^{\circ}$ and thus $\kappa_{rr} = \kappa_{\perp r}$, which allows the transport equation to be written as the following Fokker-Planck-like equation in 1D spherical coordinates \citep[e.g.,][]{Caballero-Lopez-Moraal-2004}:
\begin{equation}
%\frac{\partial f}{\partial t} = - v_{\rm{sw}} \frac{\partial f}{\partial r} + \frac{1}{r^2}\frac{\partial}{\partial r} \left(r^2 \kappa_{rr} \frac{\partial f}{\partial r}\right) + \left(\frac{\partial v_{sw}}{\partial r}\right) \frac{p}{3} \frac{\partial f}{\partial p}.
%
\frac{\partial f}{\partial t} = - v_{\rm{sw}} \frac{\partial f}{\partial r} + \frac{1}{r^2}\frac{\partial}{\partial r} \left(r^2 \kappa_{rr} \frac{\partial f}{\partial r}\right) + \frac{1}{3 r^2} \frac{\partial}{\partial r}\left( r^2 v_{\rm{sw}}\right) \frac{\partial f}{\partial \ln p}.
\end{equation}
The effective radial diffusion coefficient $\kappa_{rr}$, for $\Psi = 90^{\circ}$,  can be expressed in terms of a perpendicular diffusion mean free path as
\begin{equation}
\label{eq:kappa_rr}
\kappa_{r r} = \kappa_{\perp r} =  \frac{v \lambda_\perp}{3},
\end{equation}
with $v$ representing the particle speed. However, since in-situ observations of the corresponding parameters are not available, the form and magnitude of the diffusion parameters must be estimated based on our current understanding of these processes in the heliosphere. Although analytical forms for the perpendicular mean free path can be derived from theory, these forms require information as to the turbulence conditions in these ASPs \cite[ e.g.,][]{Engelbrecht-Burger-2013}. As such information is currently lacking, in a first approximation $\lambda_{\perp}$is modeled to scale as the inverse of the stellar magnetic field $B$, which is provided by the computations with the CRONOS code.  We, therefore, assume that
\begin{equation}
    \lambda_{\perp} = \lambda_{0} \left( \frac{B_0}{B} \right) \left( \frac{P}{P_0} \right)^{1/3}
\label{eq:lam}
\end{equation}
where $P$ is particle rigidity, $P_0 = 1$ GV, $B_0 = 1$ nT and $B$ the AMF. Furthermore, for purposes of relative comparison, $\lambda_0 = a {B_E (1~\rm{au})}/{B_1}$ is a normalized mean free path value taking into account the ratio of the stellar $B_{1}$ and solar magnetic field $B_{E}$ at 1 au as well as the ratio  $a$ of the perpendicular and parallel mean free path. As a first approach two values of this latter quantity are employed, viz. $a=0.01$ and $a=0.02$, following typical values assumed in heliospheric modulation studies \cite[ e.g.,][]{fp03}. This equation is, however, only applied inside the modulation cavity, i.e. inside the ASP. In the undisturbed ISM, the mean-free-path is chosen as $\lambda_{\perp} (r > r_{\rm{AP}}) = 1$ pc and assumed to be constant. 

Because of the stronger magnetic field in the modeled ASPs of V374 Peg and Proxima Centauri, analogously to the situation in the heliosphere, $\lambda_{\perp}$ is shorter within the cavities of these ASPs than in the ISM, where the mean free path is in the order of 0.1 to 10 pc. Cosmic-ray modulation models require essential input in some form of the local interstellar spectrum (LIS), i.e., essentially an unmodulated boundary condition. In the present study, the same LIS, namely the one used by \citet{Strauss-etal-2011}, is employed for all ASPs considered here as a first approximation. While this assumption may hold for the heliosphere and Proxima Centauri due to their relative proximity, this may not be the case for the other ASPs and will be the subject of future studies. More detailed information on the numerical solution of Fokker-Planck equations by SDEs is given, for example, in  \citet{Strauss-etal-2017} and references therein. 

\begin{figure}[!t]
\begin{center}
  \includegraphics[width=0.6\columnwidth]{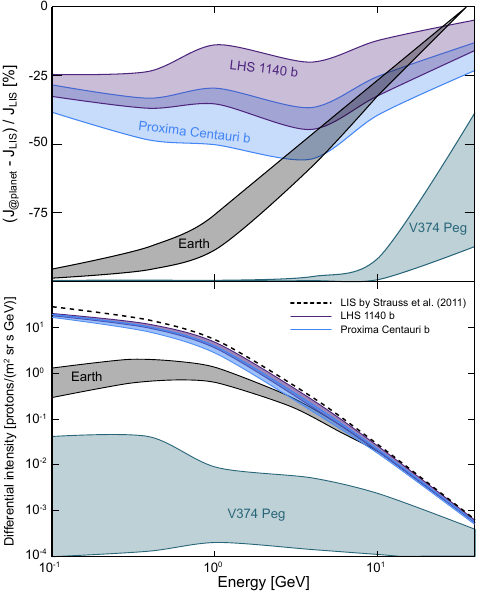}
  \caption{Upper panel: Relative change of the GCR intensity with respect to the LIS. Lower panel: Corresponding differential energy spectra within the ASPs of the heliosphere (black), V374 Peg (in petrol), Proxima Centauri (in light-blue), and LHS 1140 (in purple) at the distance of their (potential) Earth-like planets (see Sec.~\ref{sec:characteristics}).}
 \label{fig:3}
 \end{center}
 \end{figure}

The corresponding modeled primary proton energy spectra between 100~MeV and 40~GeV (lower panel) and the relative GCR modulation with respect to the unmodulated LIS are displayed in the panels of Fig.~\ref{fig:3}. Thereby, for each ASP, upper and lower limits are presented, which can be interpreted as the uncertainty due to, for example, potential stellar activity and/or possible differences in the ratio between parallel and perpendicular mean free paths \cite[e.g.,][]{EB2015}. These limits correspond to solutions computed assuming, respectively, that parameter $a$ has a value of $0.02$ or $0.01$. The corresponding energy-dependent flux values are listed in Table~\ref{tab:2}. 

As can be seen in the upper panel, the GCR flux of, for example, 1~GeV protons around Earth is reduced by about 75$\%$ to 88$\%$ of the unmodulated LIS flux during solar maximum and minimum conditions, respectively, while the flux of 1~GeV particles around V374 Peg is completely suppressed. However, in general, much less modulation of GCR protons occurs within the ASPs of Proxima Centauri and LHS 1140. Thus, in contrast to prior assumptions \citep[e.g.,][]{Sadovski-etal-2018}, the influence of GCRs on the atmospheres of the presumably Earth-like exoplanets Proxima Centauri b and LHS 1140 b is much stronger and thus cannot be neglected.

The use of a simplified 1D SDE model, however, has its limitations. For example, CR transport processes that are known to influence GCR intensities in the heliosphere and that require modeling in more than one dimension, such as drifts due to gradients and curvatures in the AMF, cannot be taken into account \cite[e.g.,][and references therein]{E17}. In the heliosphere, drift effects have long been known to significantly affect the degree to which GCR spectra are modulated \cite[e.g.,][]{jl77}. This degree would depend very much on the nature of the ASP under consideration and would need to be ascertained using 2D or 3D GCR transport models. As such, the present 1D approach serves to provide a first-order estimate of GCR modulation effects, and further refinements to this approach would be the subject of future studies. 
\section{Summary and Conclusions}
%
% 459 words
%
For the first time, we studied the ASPs of cool M-stars through 3D MHD modeling with the simulation code CRONOS. The foci of this study were the three M-stars V374 Peg, Proxima Centauri, and LHS 1140, which not only are hosts of potentially Earth-like planets in their HZs but are also rather diverse, which, for example, reflects in their different mass-loss rates, ranging from 5$\cdot 10^{-17}$ M$_\sun$ yr$^{-1}$ (LHS 1140) to 4$\cdot 10^{-10}$ M$_\sun$ yr$^{-1}$ (V374 Peg). The scenarios presented  can be seen as extreme cases of M-star ASPs. 

In contrast to previous assumptions, we found that not all M-stars drive huge ASPs: out of the three investigated ASPs, this assumption was shown only to be valid for the ASP of V374 Peg. The size of the ASP of Proxima Centauri, however, was shown to be directly comparable to the heliosphere, while the entire ASP of LHS 1140, compared to the Solar System, would fit well within the orbit of Neptune. 

Further, the influence of GCRs within the ASPs of cool stars is extensively discussed in the literature. Among others, for example, \citet{Scherer-etal-2008} studied changes in the heliospheric hydrogen flux that lead to increased fluxes of GCRs at 1~au. For the first time, this study showed the possibility of investigating the modulation of GCRs through 1D SDE modeling utilizing the results of the 3D MHD astrospheric model efforts. In contrast to the results based on analytical approaches by, for example, \citet{Sadovski-etal-2018}, which suggest that GCRs below 10 TeV are not present within M-star ASPs, we could show that the GCR flux along the line-of-sight of both the ASPs of Proxima Centauri and  LHS 1140 is up to one order of magnitude higher than within our heliosphere. The computations, although preliminarily done using a 1D GCR transport model, show that the GCR contribution cannot necessarily be assumed to be negligible. The latter, in particular, is valid for inactive cool stars, with the implication of a dominant GCR component additional to any contribution due to cosmic rays of stellar origin as discussed by, for example, \citet{Youngblood-etal-2017} and \citet{Herbst-etal-2019a}. Thus, the impact of GCRs in the context of exoplanetary habitability, in particular of the potentially Earth-like planets Proxima Centauri~b and LHS 1140~b, cannot be neglected. 

Moreover, or study points out that future studies of cool G-, K-, and M-stars will rely not only on 3D MHD modeling of their ASPs but also on 3D SDE modeling of the CR transport within. The latter is indeed possible, as the current modeling efforts can be expanded to include turbulence transport modeling, and the results can be employed in modeling more realistic GCR diffusion coefficients, as has been done in the heliosphere by, for example, \citet{tobias}. The latter will be the subject of future investigations.
%
% in total:  3490/3500 words
%
%-------------------------------------------------------------------------------------------
%-------------------------------------------------------------------------------------------
\section*{Acknowledgements}
K.H. acknowledges the International Space Science Institute and the supported International Team 464: \textit{The  Role  Of  Solar  And  Stellar  Energetic  Particles  On (Exo)Planetary Habitability (ETERNAL, \url{http://www.issibern.ch/teams/exoeternal/})}. KS is grateful to the \textit{Deut\-sche For\-schungs\-ge\-mein\-schaft (DFG)}, funding the project SCHE334/9-2. SESF acknowledges the partial financial support of the NRF under grants 93546 and 109253. This work is based on research supported partially by the CHPC in Cape Town, under the High Performance Computing project \textit{ASTRO1277}.
%---------------------------------------------------------------------------
\section*{Software}
CRONOS \citep[][]{Kissmann-etal-2018} and SDE \citep[e.g.,][]{Strauss-etal-2017}.
\bibliographystyle{aasjournal}
\bibliography{references}

\begin{thebibliography}{}
\expandafter\ifx\csname natexlab\endcsname\relax\def\natexlab#1{#1}\fi
\providecommand{\url}[1]{\href{#1}{#1}}

\bibitem[{{Anglada-Escud{\'e}} {et~al.}(2016){Anglada-Escud{\'e}}, {Amado},
  {Barnes}, {Berdi{\~n}as}, {Butler}, {Coleman}, {de La Cueva}, {Dreizler},
  {Endl}, {Giesers}, {Jeffers}, {Jenkins}, {Jones}, {Kiraga}, {K{\"u}rster},
  {L{\'o}pez-Gonz{\'a}lez}, {Marvin}, {Morales}, {Morin}, {Nelson}, {Ortiz},
  {Ofir}, {Paardekooper}, {Reiners}, {Rodr{\'\i}guez},
  {Rodr{\'\i}guez-L{\'o}pez}, {Sarmiento}, {Strachan}, {Tsapras}, {Tuomi}, \&
  {Zechmeister}}]{Anglada-Escude-etal-2016}
{Anglada-Escud{\'e}}, G., {Amado}, P.~J., {Barnes}, J., {et~al.} 2016, \nat,
  536, 437.
\newblock \url{https://doi.org/10.1038/nature19106}

\bibitem[{{Atri}(2020)}]{Atri-2020}
{Atri}, D. 2020, \mnras, 492, L28.
\newblock \url{https://doi.org/10.1093/mnrasl/slz166}

\bibitem[{{Benedict} {et~al.}(1998){Benedict}, {McArthur}, {Nelan}, {Story},
  {Whipple}, {Shelus}, {Jefferys}, {Hemenway}, {Franz}, {Wasserman},
  {Duncombe}, {van Altena}, \& {Fredrick}}]{Benedict-etal-1998}
{Benedict}, G.~F., {McArthur}, B., {Nelan}, E., {et~al.} 1998, \aj, 116, 429.
\newblock \url{https://doi.org/10.1086/300420}

\bibitem[{Bixel \& Apai(2017)}]{Bixel-Apai-2017}
Bixel, A., \& Apai, D. 2017, The Astrophysical Journal, 836, L31.
\newblock \url{https://doi.org/10.3847%2F2041-8213%2Faa5f51}

\bibitem[{{B{\"u}sching} {et~al.}(2005){B{\"u}sching}, {Kopp}, {Pohl},
  {Schlickeiser}, {Perrot}, \& {Grenier}}]{Buesching-etal-2005}
{B{\"u}sching}, I., {Kopp}, A., {Pohl}, M., {et~al.} 2005, \apj, 619, 314.
\newblock \url{https://doi.org/10.1086/426537}

\bibitem[{{Caballero-Lopez} \& {Moraal}(2004)}]{Caballero-Lopez-Moraal-2004}
{Caballero-Lopez}, R.~A., \& {Moraal}, H. 2004, Journal of Geophysical Research
  (Space Physics), 109, A01101.
\newblock \url{https://doi.org/10.1029/2003JA010098}

\bibitem[{{Candelaresi} {et~al.}(2014){Candelaresi}, {Hillier}, {Maehara},
  {Brand enburg}, \& {Shibata}}]{Candelaresi-etal-2014}
{Candelaresi}, S., {Hillier}, A., {Maehara}, H., {Brand enburg}, A., \&
  {Shibata}, K. 2014, \apj, 792, 67.
\newblock \url{https://doi.org/10.1088/0004-637X/792/1/67}

\bibitem[{{Dittmann} {et~al.}(2017){Dittmann}, {Irwin}, {Charbonneau},
  {Bonfils}, {Astudillo-Defru}, {Haywood}, {Berta-Thompson}, {Newton},
  {Rodriguez}, {Winters}, {Tan}, {Almenara}, {Bouchy}, {Delfosse}, {Forveille},
  {Lovis}, {Murgas}, {Pepe}, {Santos}, {Udry}, {W{\"u}nsche}, {Esquerdo},
  {Latham}, \& {Dressing}}]{Dittmann-etal-2017}
{Dittmann}, J.~A., {Irwin}, J.~M., {Charbonneau}, D., {et~al.} 2017, \nat, 544,
  333.
\newblock \url{https://doi.org/10.1038/nature22055}

\bibitem[{{Donati} {et~al.}(2006){Donati}, {Forveille}, {Collier Cameron},
  {Barnes}, {Delfosse}, {Jardine}, \& {Valenti}}]{Donati-etal-2006}
{Donati}, J.-F., {Forveille}, T., {Collier Cameron}, A., {et~al.} 2006,
  Science, 311, 633.
\newblock \url{https://doi.org/10.1126/science.1121102}

\bibitem[{{Engelbrecht} \& {Burger}(2013)}]{Engelbrecht-Burger-2013}
{Engelbrecht}, N.~E., \& {Burger}, R.~A. 2013, \apj, 772, 46.
\newblock \url{https://doi.org/10.1088/0004-637X/772/1/46}

\bibitem[{{Engelbrecht} \& {Burger}(2015)}]{EB2015}
---. 2015, \apj, 814, 152.
\newblock \url{https://doi.org/10.1088/0004-637X/814/2/152}

\bibitem[{{Engelbrecht} {et~al.}(2019){Engelbrecht}, {Mohlolo}, \&
  {Ferreira}}]{E17}
{Engelbrecht}, N.~E., {Mohlolo}, S.~T., \& {Ferreira}, S.~E.~S. 2019, \apjl,
  884, L54.
\newblock \url{https://doi.org/10.3847/2041-8213/ab4ad6}

\bibitem[{{Ferreira} \& {Potgieter}(2003)}]{fp03}
{Ferreira}, S.~E.~S., \& {Potgieter}, M.~S. 2003, Advances in Space Research,
  32, 657.
\newblock \url{https://doi.org/10.1016/S0273-1177(03)00360-0}

\bibitem[{{Garraffo} {et~al.}(2016){Garraffo}, {Drake}, \&
  {Cohen}}]{Garraffo-etal-2016}
{Garraffo}, C., {Drake}, J.~J., \& {Cohen}, O. 2016, \apjl, 833, L4.
\newblock \url{https://doi.org/10.3847/2041-8205/833/1/L4}

\bibitem[{{Grie{\ss}meier} {et~al.}(2016){Grie{\ss}meier}, {Tabataba-Vakili},
  {Stadelmann}, {Grenfell}, \& {Atri}}]{Griessmeier-etal-2016}
{Grie{\ss}meier}, J.~M., {Tabataba-Vakili}, F., {Stadelmann}, A., {Grenfell},
  J.~L., \& {Atri}, D. 2016, \aap, 587, A159.
\newblock \url{https://doi.org/10.1051/0004-6361/201425452}

\bibitem[{{Herbst} {et~al.}(2019{\natexlab{a}}){Herbst}, {Papaioannou},
  {Banjac}, \& {Heber}}]{Herbst-etal-2019a}
{Herbst}, K., {Papaioannou}, A., {Banjac}, S., \& {Heber}, B.
  2019{\natexlab{a}}, \aap, 621, A67.
\newblock \url{https://doi.org/10.1051/0004-6361/201832789}

\bibitem[{{Herbst} {et~al.}(2019{\natexlab{b}}){Herbst}, {Grenfell},
  {Sinnhuber}, {Rauer}, {Heber}, {Banjac}, {Scheucher}, {Schmidt}, {Gebauer},
  {Lehmann}, \& {Schreier}}]{Herbst-etal-2019b}
{Herbst}, K., {Grenfell}, J.~L., {Sinnhuber}, M., {et~al.} 2019{\natexlab{b}},
  \aap, 631, A101.
\newblock \url{https://doi.org/10.1051/0004-6361/201935888}

\bibitem[{{Izmodenov} \& {Alexashov}(2015)}]{Izmodenov-Alexashov-2015}
{Izmodenov}, V.~V., \& {Alexashov}, D.~B. 2015, \apjs, 220, 32.
\newblock \url{https://doi.org/10.1088/0067-0049/220/2/32}

\bibitem[{{Jokipii} \& {Levy}(1977)}]{jl77}
{Jokipii}, J.~R., \& {Levy}, E.~H. 1977, \apjl, 213, L85.
\newblock \url{https://doi.org/10.1086/182415}

\bibitem[{{Kay} {et~al.}(2016){Kay}, {Opher}, \& {Kornbleuth}}]{Kay-etal-2016}
{Kay}, C., {Opher}, M., \& {Kornbleuth}, M. 2016, \apj, 826, 195.
\newblock \url{https://doi.org/10.3847/0004-637X/826/2/195}

\bibitem[{{Khodachenko} {et~al.}(2007){Khodachenko}, {Ribas}, {Lammer},
  {Grie{\ss}meier}, {Leitner}, {Selsis}, {Eiroa}, {Hanslmeier}, {Biernat},
  {Farrugia}, \& {Rucker}}]{Khodachenko-etal-2007}
{Khodachenko}, M.~L., {Ribas}, I., {Lammer}, H., {et~al.} 2007, Astrobiology,
  7, 167

\bibitem[{{Kissmann} {et~al.}(2018){Kissmann}, {Kleimann}, {Krebl}, \&
  {Wiengarten}}]{Kissmann-etal-2018}
{Kissmann}, R., {Kleimann}, J., {Krebl}, B., \& {Wiengarten}, T. 2018, \apjs,
  236, 53.
\newblock \url{https://doi.org/10.3847/1538-4365/aabe75}

\bibitem[{{Lammer} {et~al.}(2007){Lammer}, {Lichtenegger}, {Kulikov},
  {Grie{\ss}meier}, {Terada}, {Erkaev}, {Biernat}, {Khodachenko}, {Ribas},
  {Penz}, \& {Selsis}}]{Lammer-etal-2007}
{Lammer}, H., {Lichtenegger}, H. I.~M., {Kulikov}, Y.~N., {et~al.} 2007,
  Astrobiology, 7, 185.
\newblock \url{https://doi.org/10.1089/ast.2006.0128}

\bibitem[{{Mohanty} {et~al.}(2002){Mohanty}, {Basri}, {Shu}, {Allard}, \&
  {Chabrier}}]{Mohanty-etal-2002}
{Mohanty}, S., {Basri}, G., {Shu}, F., {Allard}, F., \& {Chabrier}, G. 2002,
  \apj, 571, 469.
\newblock \url{https://doi.org/10.1086/339911}

\bibitem[{{Morin} {et~al.}(2008){Morin}, {Donati}, {Forveille}, {Delfosse},
  {Dobler}, {Petit}, {Jardine}, {Collier Cameron}, {Albert}, {Manset},
  {Dintrans}, {Chabrier}, \& {Valenti}}]{Morin-etal-2008}
{Morin}, J., {Donati}, J.~F., {Forveille}, T., {et~al.} 2008, \mnras, 384, 77.
\newblock \url{https://doi.org/10.1111/j.1365-2966.2007.12709.x}

\bibitem[{Moschou {et~al.}(2019)Moschou, Drake, Cohen, Alvarado-G{\'{o}}mez,
  Garraffo, \& Fraschetti}]{Moschou-etal-2019}
Moschou, S.-P., Drake, J.~J., Cohen, O., {et~al.} 2019, The Astrophysical
  Journal, 877, 105.
\newblock \url{https://doi.org/10.3847%2F1538-4357%2Fab1b37}

\bibitem[{{M{\"u}ller} {et~al.}(2006){M{\"u}ller}, {Frisch}, {Florinski}, \&
  {Zank}}]{Mueller-etal-2006}
{M{\"u}ller}, H.-R., {Frisch}, P.~C., {Florinski}, V., \& {Zank}, G.~P. 2006,
  \apj, 647, 1491.
\newblock \url{https://doi.org/10.1086/505588}

\bibitem[{{Parker}(1958)}]{Parker-1958}
{Parker}, E.~N. 1958, \apj, 128, 664.
\newblock \url{https://doi.org/10.1086/146579}

\bibitem[{{Parker}(1965)}]{Parker-1965}
---. 1965, \planss, 13, 9.
\newblock \url{https://doi.org/10.1016/0032-0633(65)90131-5}

\bibitem[{{Preusse} {et~al.}(2005){Preusse}, {Kopp}, {B{\"u}chner}, \&
  {Motschmann}}]{Preusse-etal-2005}
{Preusse}, S., {Kopp}, A., {B{\"u}chner}, J., \& {Motschmann}, U. 2005, \aap,
  434, 1191.
\newblock \url{https://doi.org/10.1051/0004-6361:20041680}

\bibitem[{{Reiners} \& {Basri}(2008)}]{Reiners-Basri-2008}
{Reiners}, A., \& {Basri}, G. 2008, \aap, 489, L45.
\newblock \url{https://doi.org/10.1051/0004-6361:200810491}

\bibitem[{{Sadovski} {et~al.}(2018){Sadovski}, {Struminsky}, \&
  {Belov}}]{Sadovski-etal-2018}
{Sadovski}, A.~M., {Struminsky}, A.~B., \& {Belov}, A. 2018, Astronomy Letters,
  44, 324.
\newblock \url{https://doi.org/10.1134/S1063773718040072}

\bibitem[{{Scherer} {et~al.}(2020){Scherer}, {Baalmann}, {Fichtner},
  {Kleimann}, {Bomans}, {Weis}, {Ferreira}, \& {Herbst}}]{Scherer-etal-2020}
{Scherer}, K., {Baalmann}, L.~R., {Fichtner}, H., {et~al.} 2020, \mnras, 493,
  4172.
\newblock \url{https://doi.org/10.1093/mnras/staa497}

\bibitem[{{Scherer} {et~al.}(2014){Scherer}, {Fichtner}, {Fahr}, {Bzowski}, \&
  {Ferreira}}]{Scherer-etal-2014}
{Scherer}, K., {Fichtner}, H., {Fahr}, H.~J., {Bzowski}, M., \& {Ferreira},
  S.~E.~S. 2014, \aap, 563, A69.
\newblock \url{https://doi.org/10.1051/0004-6361/201321151}

\bibitem[{{Scherer} {et~al.}(2008){Scherer}, {Fichtner}, {Heber}, {Ferreira},
  \& {Potgieter}}]{Scherer-etal-2008}
{Scherer}, K., {Fichtner}, H., {Heber}, B., {Ferreira}, S.~E.~S., \&
  {Potgieter}, M.~S. 2008, Advances in Space Research, 41, 1171.
\newblock \url{https://doi.org/10.1016/j.asr.2007.03.016}

\bibitem[{{Scherer} {et~al.}(2015){Scherer}, {van der Schyff}, {Bomans},
  {Ferreira}, {Fichtner}, {Kleimann}, {Strauss}, {Weis}, {Wiengarten}, \&
  {Wodzinski}}]{Scherer-etal-2015}
{Scherer}, K., {van der Schyff}, A., {Bomans}, D.~J., {et~al.} 2015, \aap, 576,
  A97.
\newblock \url{https://doi.org/10.1051/0004-6361/201425091}

\bibitem[{{Scheucher} {et~al.}(2018){Scheucher}, {Grenfell}, {Wunderlich},
  {Godolt}, {Schreier}, \& {Rauer}}]{Scheucher-etal-2018}
{Scheucher}, M., {Grenfell}, J.~L., {Wunderlich}, F., {et~al.} 2018, \apj, 863,
  6.
\newblock \url{https://doi.org/10.3847/1538-4357/aacf03}

\bibitem[{Scheucher {et~al.}(2020)Scheucher, Herbst, Schmidt, Grenfell,
  Schreier, Banjac, Heber, Rauer, \& Sinnhuber}]{Scheucher-etal-2020a}
Scheucher, M., Herbst, K., Schmidt, V., {et~al.} 2020, The Astrophysical
  Journal, 893, 12.
\newblock \url{https://doi.org/10.3847%2F1538-4357%2Fab7b74}

\bibitem[{{Smith} \& {Scalo}(2009)}]{Smith-Scalo-2009}
{Smith}, D.~S., \& {Scalo}, J.~M. 2009, Astrobiology, 9, 673.
\newblock \url{https://doi.org/10.1089/ast.2009.0337}

\bibitem[{{Strauss} {et~al.}(2011){Strauss}, {Potgieter}, {B{\"u}sching}, \&
  {Kopp}}]{Strauss-etal-2011}
{Strauss}, R.~D., {Potgieter}, M.~S., {B{\"u}sching}, I., \& {Kopp}, A. 2011,
  \apj, 735, 83.
\newblock \url{https://doi.org/10.1088/0004-637X/735/2/83}

\bibitem[{{Strauss} \& {Effenberger}(2017)}]{Strauss-etal-2017}
{Strauss}, R. D.~T., \& {Effenberger}, F. 2017, \ssr, 212, 151.
\newblock \url{https://doi.org/10.1007/s11214-017-0351-y}

\bibitem[{{van Marle} {et~al.}(2014){van Marle}, {Cox}, \&
  {Decin}}]{vanMarle-etal-2014}
{van Marle}, A.~J., {Cox}, N.~L.~J., \& {Decin}, L. 2014, \aap, 570, A131.
\newblock \url{https://doi.org/10.1051/0004-6361/201424452}

\bibitem[{{Vida} {et~al.}(2016){Vida}, {Kriskovics}, {Ol{\'a}h}, {Leitzinger},
  {Odert}, {K{\H{o}}v{\'a}ri}, {Korhonen}, {Greimel}, {Robb}, {Cs{\'a}k}, \&
  {Kov{\'a}cs}}]{Vida-etal-2016}
{Vida}, K., {Kriskovics}, L., {Ol{\'a}h}, K., {et~al.} 2016, \aap, 590, A11.
\newblock \url{https://doi.org/10.1051/0004-6361/201527925}

\bibitem[{{Vidotto} {et~al.}(2011){Vidotto}, {Jardine}, {Opher}, {Donati}, \&
  {Gombosi}}]{Vidotto-etal-2011}
{Vidotto}, A.~A., {Jardine}, M., {Opher}, M., {Donati}, J.~F., \& {Gombosi},
  T.~I. 2011, \mnras, 412, 351.
\newblock \url{https://doi.org/10.1111/j.1365-2966.2010.17908.x}

\bibitem[{{West} {et~al.}(2015){West}, {Weisenburger}, {Irwin},
  {Berta-Thompson}, {Charbonneau}, {Dittmann}, \& {Pineda}}]{West-etal-2015}
{West}, A.~A., {Weisenburger}, K.~L., {Irwin}, J., {et~al.} 2015, \apj, 812, 3.
\newblock \url{https://doi.org/10.1088/0004-637X/812/1/3}

\bibitem[{{West} {et~al.}(2004){West}, {Hawley}, {Walkowicz}, {Covey},
  {Silvestri}, {Raymond}, {Harris}, {Munn}, {McGehee}, {Ivezi{\'c}}, \&
  {Brinkmann}}]{West-etal-2004}
{West}, A.~A., {Hawley}, S.~L., {Walkowicz}, L.~M., {et~al.} 2004, \aj, 128,
  426.
\newblock \url{https://doi.org/10.1086/421364}

\bibitem[{{Wiengarten} {et~al.}(2016){Wiengarten}, {Oughton}, {Engelbrecht},
  {Fichtner}, {Kleimann}, \& {Scherer}}]{tobias}
{Wiengarten}, T., {Oughton}, S., {Engelbrecht}, N.~E., {et~al.} 2016, \apj,
  833, 17.
\newblock \url{https://doi.org/10.3847/0004-637X/833/1/17}

\bibitem[{{Wilkin}(2000)}]{Wilkin-2000}
{Wilkin}, F.~P. 2000, \apj, 532, 400.
\newblock \url{https://doi.org/10.1086/308576}

\bibitem[{{Wood} {et~al.}(2001){Wood}, {Linsky}, {M{\"u}ller}, \&
  {Zank}}]{Wood-etal-2001}
{Wood}, B.~E., {Linsky}, J.~L., {M{\"u}ller}, H.-R., \& {Zank}, G.~P. 2001,
  \apjl, 547, L49.
\newblock \url{https://doi.org/10.1086/318888}

\bibitem[{{Youngblood} {et~al.}(2017){Youngblood}, {France}, {Loyd}, {Brown},
  {Mason}, {Schneider}, {Tilley}, {Berta-Thompson}, {Buccino}, {Froning},
  {Hawley}, {Linsky}, {Mauas}, {Redfield}, {Kowalski}, {Miguel}, {Newton},
  {Rugheimer}, {Segura}, {Roberge}, \& {Vieytes}}]{Youngblood-etal-2017}
{Youngblood}, A., {France}, K., {Loyd}, R.~O.~P., {et~al.} 2017, \apj, 843, 31.
\newblock \url{https://doi.org/10.3847/1538-4357/aa76dd}

\end{thebibliography}

\end{document}